\documentclass[journal,twocolumn,twoside]{IEEEtranTCOM}
\normalsize

\usepackage{graphicx}
\usepackage{epsfig}
\usepackage{psfrag}
\usepackage{color}
\usepackage{amsmath}
\usepackage{amssymb}
\usepackage{eso-pic}

\usepackage{cite}
\usepackage{dsfont}
\usepackage{epstopdf}

\usepackage[hypertex]{hyperref}

\begin{document}
\title{New Signal Designs for Enhanced Spatial Modulation}
\author{Chien-Chun~Cheng,~\IEEEmembership{Student Member,~IEEE}, Hikmet~Sari,~\IEEEmembership{Fellow,~IEEE},\\
Serdar Sezginer,~\IEEEmembership{Senior Member,~IEEE}, and Yu T.
Su,~\IEEEmembership{Senior Member,~IEEE}
\thanks{H. Sari and C. C. Cheng are with the Department of
Telecommunications, CentraleSup\'{e}lec, Plateau de Moulon, 1-3 Rue Joliot
Curie, 91190 Gif sur Yvette, France. (E-mail: hikmet.sari@supelec.fr, chien-chun.cheng@supelec.fr)}
\thanks{S. Sezginer and H. Sari are with SEQUANS
Communications, Les Portes de La Defense, 15-55, Boulevard Charles de Gaulle, 92700, Colombes, France
(e-mail: serdar@sequans.com, hikmet@sequans.com)}
\thanks{Y. T. Su and C. C. Cheng are with the Department of
Electrical Engineering, National Chiao Tung University, Hsinchu,
30056, Taiwan (e-mail: cccheng.cm98g@nctu.edu.tw,
ytsu@mail.nctu.edu.tw)}
}

\markboth{Submitted to IEEE Transactions on Wireless Communications}
{Submitted paper}
\maketitle

%
%
\begin{abstract}
In this paper, we present three new signal designs for Enhanced Spatial Modulation (ESM), which was recently introduced by the present authors. The basic idea of ESM is to convey information bits not only by the index(es) of the active transmit antenna(s) as in conventional Spatial Modulation (SM), but also by the types of the signal constellations used. The original ESM schemes were designed with reference to single-stream SM and involved one or more secondary modulations in addition to the primary modulation. Compared to single-stream SM, they provided either higher throughput or improved signal-to-noise ratio (SNR). In the present paper, we focus on multi-stream SM (MSM) and present three new ESM designs leading to increasing SNR gains when they are operated at the same spectral efficiency. The secondary signal constellations used in the first two designs are based on a single geometric interpolation step in the signal constellation plane, while the third design also makes use of additional constellations derived through a second interpolation step. The new ESM signal designs are described for MIMO systems with four transmit antennas two of which are active, but we also briefly present extensions to higher numbers of antennas. Theoretical analysis and simulation results indicate that the proposed designs provide a significant SNR gain over MSM.
\end{abstract}

\IEEEpeerreviewmaketitle


%
%
\section{Introduction}
Multiple-input multiple-output (MIMO) technologies are now widely used in wireless communications systems standards. The objective of using these technologies is to increase data throughput, increase performance, and make different trade-offs between these two desired features. The main problem limiting the practical implementation of MIMO technologies is related to the decoding complexity, which increases with the number of antennas \cite{fun_05,wc_05,mul_09}. In a number of cases, cost and energy consumption considerations lead to the implementation of a smaller number of radio-frequency (RF) chains in the transmitter than the number of transmit antennas. This is often the case in mobile and fixed user equipment, because the number of antennas is typically dictated by the performance requirements for the downlink signal, and cost and energy consumption limitations may not allow the implementation of as many RF chains. Spatial modulation (SM) is a MIMO technique, which was precisely introduced for those cases. 

The first papers on SM considered MIMO systems with a single transmit RF chain \cite{spa_08v,spa_11,sin_11,spa_01}. These SM schemes convey information bits by allocating them to the active antenna index, while transmitting a group of other bits through the symbols transmitted from the selected active antenna. Further work on SM generalized this technique by relaxing the single transmit RF-chain constraint and allowing more than one antenna to transmit simultaneously, see e.g., \cite{gen_10g,gen_10,gen_12}. A comprehensive survey on Generalized SM appears in \cite{spa_14}. A simple variant of SM is the so-called Space-Shift Keying (SSK) \cite{spa_09}, where only the index of the active antenna transmits information. In other words, the active antenna in SSK does not transmit any data symbols, but instead an unmodulated signal. This concept too was naturally extended to multiple active antennas \cite{gen_08}, and the resulting scheme was coined Generalized SSK. The literature on SM, SSK, and their generalized versions is now quite abundant; we mention here \cite{coh_10,stb_11,gen_11,spa_14c}, which address space-time code design, and \cite{per_13,gen_13s}, which address the decoding aspects.

But even in its multi-stream version, the spectral efficiency of SM remains modest compared to spatial multiplexing (SMX) \cite{fun_05}, which is widely used in conventional MIMO systems. In order to improve spectral efficiency, the present authors recently introduced a new SM concept in \cite{esm_15} using multiple signal constellations. This technique, referred to as Enhanced SM (ESM), conveys information bits using one or two active transmit antennas and two or more reduced-size secondary modulations in addition to the primary modulation. The primary modulation in that scheme was restricted to the periods of one active antenna, and the secondary modulations were used with two active transmit antennas. A significant performance gain was achieved compared to conventional SM when the two techniques are operated at the same spectral efficiency. In the comparisons, conventional SM employed one active transmit (TX) antenna only, because the ESM design of \cite{esm_15} was made with reference to single-stream SM. 

In this paper, we introduce three new ESM designs taking as reference Multi-stream SM (MSM) \cite{per_13}. The description is made for MIMO systems with four transmit antennas two of which are active, but generalization to higher numbers of antennas is also briefly presented. As in \cite{esm_15}, the basic principle is to use additional modulations with the primary modulation in order to increase the number of antenna and modulation combinations. The first two ESM schemes use a secondary constellation that is derived through a single-step geometric interpolation between the primary constellation points. When the indexes of two active TX antennas are selected, the first scheme transmits a symbol from the primary constellation on one antenna and a symbol from the secondary constellation on the other. The second type of ESM does not use the primary signal constellation in full. Instead, it uses subsets in such a way as to further reduce the average transmit energy. The third ESM scheme introduces a second step of geometric interpolation, which leads to the derivation of two additional constellations. The signal space is constructed over blocks of two consecutive channel uses in order to preserve the minimum Euclidean distance despite the reduced distance between the different constellations used. The mathematical analysis and the simulation results indicate that the proposed schemes provide a significant performance gain with respect to MSM. Parts of this work were presented in \cite{esm_15p}.

The paper is organized as follows: In Section II, we give a brief description of the system model and formulate the ESM design problem. In Section III, we present a brief review of MSM and describe the proposed ESM designs for MIMO systems with four transmit antennas (4-TX) and M-QAM as primary modulation. In Section IV, we extend our designs to MIMO systems with a higher number of antennas. Error rate performance and receiver complexity are investigated in Section V. Finally, the simulation results are reported in Section VI, and our conclusions are given in Section VII.

%
%
\section{System Model and Problem Formulation}
For a MIMO system operating on Rayleigh fading channels, the received signal can be expressed as:
%
%
\begin{align}
\mathbf{y} = \mathbf{H} \mathbf{x} + \mathbf{n}, \label{eq.system_model}
\end{align}
where $N_R$ denotes the number of receive antennas, $N_T$ is the number of transmit antennas, $\mathbf{H}$ is the $N_R \times N_T$ channel matrix, $\mathbf{x}$ is the $N_T \times 1$ transmitted symbol vector, and $\mathbf{n}$ designates the additive white Gaussian noise (AWGN). Assume that the entries of the channel matrix $\textbf{H}$ are independent and identically distributed (i. i. d.) complex circularly symmetric Gaussian variables of the form $\mathcal{N}_{c}(0,1)$ and the entries of AWGN, $\mathbf{n}$, are i. i. d. Gaussian noise of the form  $\mathcal{N}_{c}(0,N_0)$. The transmit energy is $\mathbb{E}[\mathbf{x}^H\mathbf{x}] = E_s$, and the average signal-to-noise ratio (SNR) is defined as $\mbox{SNR} = E_s/N_0$. Note that the main difference between SM and conventional MIMO is that in the former not all transmit antennas are activated simultaneously, which means that there are some zero elements in the transmit symbol vector $\mathbf{x}$. When only two transmit antennas are active, a convenient representation of the transmitted codeword $\mathbf{x}$ is as follows:
%
\begin{align}
\mathbf{x} =  \begin{bmatrix} 0,\cdots,0,x_m,0,\cdots,0,x_n, 0,\cdots,0 \end{bmatrix}^T, \label{eq.sm_x}
\end{align}
for $m\neq n$. Here vector $\textbf{x}$ is of dimension $N_T$, $m$ and $n$ with $m=1,\cdots, N_T$, and $n=1,\cdots,N_T$ are the indexes of the two active TX antennas, and $x_m$ and $x_n$ denote the symbols transmitted from these two antennas. This representation is easily generalized to cases with a higher number of active antennas by introducing in (\ref{eq.sm_x}) as many non-zero components as the number of active antennas $N_A$. The new ESM designs will be introduced in the next section for $N_T=4$ and $N_A=2$. Generalization of this ESM concept to higher number of  transmit and active antennas will be described in Section IV.   

As in \cite{esm_15}, we use here the concept of multiple constellations in order to increase the number of codewords beyond that given by the indexes of the active transmit antennas and the primary constellation alone. The basic principle of our design is to preserve in the signal space the minimum Euclidean distance $\delta_0$ of the primary constellation. The additional constellations too have a minimum Euclidean distance of $\delta_0$, but the minimum distance between points selected from different constellations is smaller than this value. Note that the additional constellations are derived using optimum geometric interpolation in the primary constellation plane, which consists of placing the points of these constellations at the centers of the squares formed by neighbor points of M-QAM used as primary constellation.  This choice guarantees a minimum distance of ${\delta_0}/{\sqrt{2}}$ between the points of the primary constellation and those of the secondary constellation derived after the first interpolation step. Similarly, it guarantees a minimum distance of ${\delta_0}/{2}$ between the points of the primary and secondary constellations and those of the third and the fourth constellations derived after the second interpolation step. Using these multiple constellations, a minimum Euclidean distance of $\delta_0$ is preserved in the signal space by imposing that codewords differ in two or more components depending on the constellations from which the non-zero components take their values. 

%
%
\section{Enhanced SM (ESM)}
Before introducing our proposed ESM designs, we first briefly describe the baseline Multi-stream SM (MSM) scheme \cite{per_13}, which will be used as basis for comparisons.

\subsection{Baseline: MSM}
MSM with four TX antennas ($N_T = 4$) out of which two are active ($N_A = 2$) and transmitting M-QAM symbols can be described using the following signal space  representation:
%
%
\begin{align}
\mathbf{x} \in 
\left \{  \begin{smallmatrix}
\begin{bmatrix} P_{M} \\ P_{M}\\0\\0 \end{bmatrix},  
\begin{bmatrix} 0 \\ 0 \\P_{M}\\P_{M} \end{bmatrix},
\begin{bmatrix} P_{M}\\0\\P_{M}\\0 \end{bmatrix},
\begin{bmatrix} 0 \\P_{M}\\0\\P_{M} \end{bmatrix} \end{smallmatrix}\right \}, \label{eq.msm}
\end{align}   
where the entry $P_{M}$ denotes the M-QAM constellation, and the zero entries correspond to the silent transmit antennas. This MSM scheme achieves a throughput of $2 + 2\log_2 M$ bits per channel use (bpcu). Indeed, $2$ information bits are assigned to select one of the four active antenna combinations which appear in (\ref{eq.msm}), and $2\log_2 M$ bits select two symbols from the $P_M$ signal constellation to be transmitted from the two active antennas. The throughput is $10$ bpcu with 16QAM and $14$ bpcu with 64QAM. The total energy per transmitted codeword is $E_s   = 20$ for 16QAM and $E_s   = 84$ for 64QAM, which is twice the average symbol energy.

We now describe our first ESM design, which we refer to as ESM-Type1 in the sequel.

%
%
\subsection{ESM-Type1}
\begin{figure}[!t]
\centering
\includegraphics[width=2.5in,height=2in]{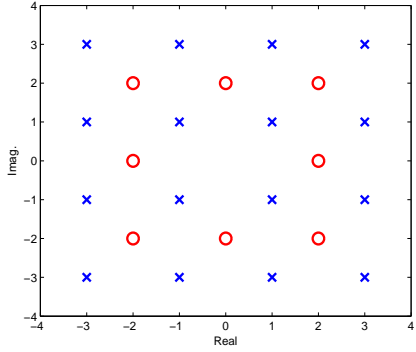}
\caption{The constellations used in ESM-Type1 with $M = 16$. The blue crosses represent 16QAM, and the red circles represent constellation $S_8$.}
\label{F.es16}
\end{figure}

\begin{figure}[!t]
\centering
\includegraphics[width=2.5in,height=2in]{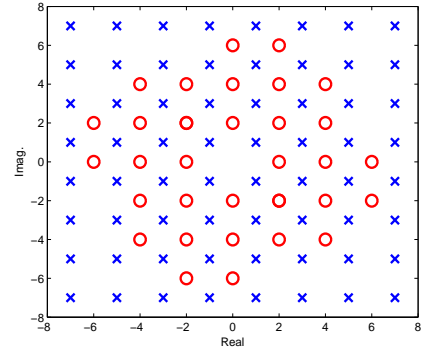}
\caption{The constellations used in ESM-Type1 with $M = 64$. The blue crosses represent 64QAM, and the red circles represent $S_{32}$.}
\label{F.esm-type1-64qam}
\end{figure}

For the same spectral efficiency as the MSM scheme described above, the transmitted codeword $\mathbf{x}$ in this design are given by:
%
%
\begin{align}
\mathbf{x} \in \left\{ \begin{smallmatrix}
\begin{bmatrix} P_{M} \\ S_{{M}/{2}}\\0\\0 \end{bmatrix},  
\begin{bmatrix} P_{M}  \\ 0 \\ 0 \\S_{{M}/{2}} \end{bmatrix},
\begin{bmatrix} 0\\P_{M} \\S_{{M}/{2}}\\0 \end{bmatrix},
\begin{bmatrix} 0 \\0\\P_{M}\\S_{{M}/{2}} \end{bmatrix} 
\\
\\
\begin{bmatrix} S_{{M}/{2}} \\ P_{M}\\0\\0 \end{bmatrix},  
\begin{bmatrix} S_{{M}/{2}}  \\ 0 \\ 0 \\P_{M} \end{bmatrix},
\begin{bmatrix} 0\\S_{{M}/{2}} \\P_{M}\\0 \end{bmatrix},
\begin{bmatrix} 0 \\0\\S_{{M}/{2}}\\P_{M} \end{bmatrix} 
\end{smallmatrix} \right\}, \label{eq.esmtype1}
\end{align}
Here, we have $8$ antenna and constellation combinations in the signal space: As in MSM, there are four active antenna combinations, but while one of the active antennas transmits a symbol from the primary M-QAM constellation $P_M$, the other antenna transmits a symbol from a secondary constellation of half size, referred to as $S_{{M}/{2}}$. The two signal constellations are shown in Fig. \ref{F.es16} for $M = 16$ and in Fig. \ref{F.esm-type1-64qam} for $M = 64$. The secondary signal constellation $S_{{M}/{2}}$ has the following mathematical representation for $M = 16$ and $M = 64$: 
%
%
\begin{align}
S_8 =\{ \pm 2\pm 2i, \pm 2, \pm 2i \}, \notag
\end{align}
and
%
%
\begin{align}
S_{32} =  \left\{  
				 \begin{matrix} 
				 S_8, \pm 4,\pm 4i,\pm 6,\pm 6i \\
                 \pm 4 \pm 2i,\pm4\pm4i,\pm 2 \pm 4i\\
                 2+6i, 6-2i,-6+2i,-2-6i 
                 \end{matrix} 
                 \right\}. \notag
\end{align}
Similarly to the baseline MSM of the previous subsection, this ESM design achieves a throughput of $2 + 2 \log_2 M$ bpcu despite the fact that one of the antennas transmits symbols from a half-size signal constellation. Indeed, the two symbols transmitted in parallel from the two active TX antennas convey $2\log_2 M -1$ bits only, but the number of antenna/constellation combinations (pairs of $m$, $n$ indexes along with the assigned signal constellations) is $8$ in this case, and therefore $3$ bits must be assigned to select one of these combinations.

%
%

%
%

Let us now examine the average energy per transmitted codeword. To evaluate the average codeword energy, we first need to evaluate the average energy of the secondary constellation used. A simple inspection of Figs. \ref{F.es16} and \ref{F.esm-type1-64qam} indicates that the average energy per symbol is $E_{S_8}=6$ for the $S_8$ constellation and $E_{S_{32}}=22$ for the $S_{32}$ constellation. Since the average energy per $16$QAM symbol is $E_{16QAM} = 10$ and the average energy per 64QAM symbol is $E_{64QAM}=42$, the average energy per transmitted codeword in this scheme is $16$ for $M = 16$ and $64$ for $M = 64$. We summarize these properties as follows: 
%
%
\begin{align}
E_{ESM-Type1-16QAM} = 16, \notag
\end{align}
and
%
%
\begin{align}
E_{ESM-Type1-64QAM} = 64. \notag
\end{align}
This means that in terms of total transmit energy, ESM-Type1 with 16QAM (resp. 64QAM) as primary modulation saves 20\% (resp. 24\%) compared to baseline MSM. In the dB scale, this corresponds to a gain of 1 dB (resp. 1.2 dB).

%
%
\subsection{ESM-Type2}
By using an $M$-point primary constellation and a half-size secondary constellation (with ${M}/{2}$ points), we managed to reduce the total transmit energy to some extent using ESM-Type1. We will now describe a second ESM design, which brings additional gain. In this design, which we refer to as ESM-Type2, we do not use the original primary constellation $P_M$ in full, but instead a subset $P_{{M}/{2}}$, which consists of the ${M}/{2}$ points of smallest energy. For $M = 64$, $P_{{M}/{2}}$ is the conventional 32QAM signal constellation, and for $M = 16$, it is a (non-conventional) 8QAM signal constellation given by: 
%
%
\begin{align}
P_8 = \{ \pm 1\pm i,3+i, 1-3i,-3-i,-1+3i \}. \notag
\end{align}
In ESM-Type2, the design procedure is as follows: The transmitted codewords $\mathbf{x}$ belong to a signal space $L$, which is the union of four subspaces $L_1$, $L_2$, $L_3$, $L_4$: 
%
%
\begin{align}
\textbf{x} \in \{ L_1, L_2, L_3, L_4 \}. \label{ESM_ty2}
\end{align}
The first three subspaces are defined as:
%
%
\begin{align}
& L_1 = \left\{ \begin{smallmatrix}
\begin{bmatrix} P_{M/2} \\ S_{M/2}\\0\\0 \end{bmatrix},  
\begin{bmatrix} S_{M/2}  \\ P_{M/2} \\ 0 \\0 \end{bmatrix},
\begin{bmatrix} 0\\0 \\P_{M/2}\\S_{M/2} \end{bmatrix},
\begin{bmatrix} 0 \\0\\S_{M/2}\\P_{M/2} \end{bmatrix}  \end{smallmatrix} \right\}, \label{ty2_L1} \\
& L_2 = \left\{ \begin{smallmatrix}
\begin{bmatrix} P_{M/2} \\0\\ S_{M/2} \\ 0 \end{bmatrix},  
\begin{bmatrix} S_{M/2}  \\ 0 \\ P_{M/2} \\ 0 \end{bmatrix},
\begin{bmatrix} 0 \\P_{M/2} \\ 0 \\S_{M/2} \end{bmatrix},
\begin{bmatrix} 0 \\ S_{M/2} \\0 \\P_{M/2} \end{bmatrix}  \end{smallmatrix} \right\}, \label{ty2_L2}\\
& L_3 = \left\{ \begin{smallmatrix}
\begin{bmatrix} P_{M/2} \\ 0\\0\\ S_{M/2} \end{bmatrix},  
\begin{bmatrix} S_{M/2}  \\ 0 \\ 0 \\ P_{M/2} \end{bmatrix},
\begin{bmatrix} 0 \\P_{M/2}\\S_{M/2} \\0 \end{bmatrix},
\begin{bmatrix} 0 \\ S_{M/2}\\P_{M/2} \\0 \end{bmatrix}  \end{smallmatrix} \right\}. \label{ty2_L3} 
\end{align}
Different subspaces use different active antenna combinations, but in all of these three subspaces one active antenna transmits symbols from the $P_{M/2}$ signal constellation, while the other active antenna transmits symbols from the $S_{M/2}$ constellation. Note that $2\log_2M - 2$ information bits are conveyed by the transmitted symbols, and 2 information bits are used to select one antenna combination in each subspace. Also, 2 prefix bits select a particular $L_j$ subspace, and hence the total number of bits per channel use is $2+2\log_2 M$. 

%
%
\begin{figure}[!t]
\centering
\includegraphics[width=2.5in,height=2in]{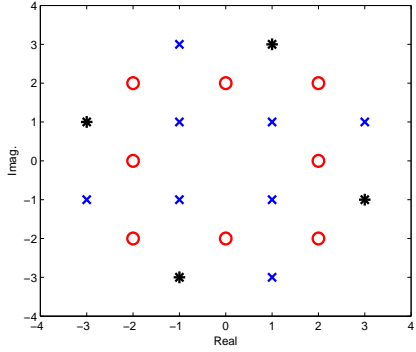}
\caption{The constellations used in ESM-Type2 with $M = 16$. The blue crosses represent $P_8$, the red circles represent $S_8$, and the black stars represent $Q_4$.}
\label{F.esm-type2}
\end{figure}

The fourth signal subspace $L_4$ is more involved. For $M = 16$, it is given by
%
%
\begin{align}
L_4 = \left\{  \begin{smallmatrix} 
\begin{bmatrix} Q_{4} \\ S_{8}\\0\\0 \end{bmatrix},  
\begin{bmatrix} S_{8}  \\ Q_{4} \\ 0 \\0 \end{bmatrix},
\begin{bmatrix} 0\\0 \\Q_{4}\\S_8 \end{bmatrix},
\begin{bmatrix} 0 \\0\\S_{8}\\Q_{4} \end{bmatrix} 
\\
\\  
\begin{bmatrix} Q_{4} \\ 0\\ S_{8}\\0 \end{bmatrix},  
\begin{bmatrix} S_{8}  \\ 0\\ Q_{4} \\0 \end{bmatrix},
\begin{bmatrix} 0\\Q_{4}\\0\\S_8 \end{bmatrix},
\begin{bmatrix} 0 \\S_{8}\\0\\Q_{4} \end{bmatrix} 
 \end{smallmatrix}  \right\}. \label{eq_esm_ty2_L4}
\end{align}
In this subspace, one of the active antennas transmits a symbol from the $S_8$ constellation, while the other antenna transmits a symbol from a $Q_4$ signal constellation, defined as follows: 
\begin{align}
Q_4 = \begin{Bmatrix} 1+3i,3-i, -1-3i, -3+i
\end{Bmatrix}. \notag
\end{align}
This constellation is shown in Fig. \ref{F.esm-type2} together with $P_8$ and $S_8$. The symbols in signal subspace $L_4$ carry $5$ information bits only, but this subspace includes $8$ active antenna and modulation combinations, and therefore $3$ bits are needed to select one of them. Together with the prefix bits assigned to the $L_4$ subspace itself, $10$ bits are transmitted per each channel use.

For $M = 64$, direct extension of the $L_4$ subspace as given by (\ref{eq_esm_ty2_L4}) is not optimal in terms of transmit energy. Direct extension means that the $S_8$ and the $Q_4$ constellations in (\ref{eq_esm_ty2_L4}) are replaced by $S_{32}$ and $Q_{16}$, where $Q_{16}$ consists of a $16$-point extension of  $S_{32}$. Instead, we found that the following choice of subspace $L_4$ minimizes the average transmit energy:
\begin{align}
L_4 = \begin{Bmatrix} L_5, L_6  \end{Bmatrix}
\end{align}
with
%
%
\begin{align}
L_5 = \left\{  \begin{smallmatrix} 
\begin{bmatrix} Q_{8} \\ P_{32}\\0\\0 \end{bmatrix},  
\begin{bmatrix} P_{32}  \\ Q_{8} \\ 0 \\0 \end{bmatrix},
\begin{bmatrix} 0\\0 \\Q_{8}\\P_{32} \end{bmatrix},
\begin{bmatrix} 0 \\0\\P_{32}\\Q_{8} \end{bmatrix} 
\\
\\  
\begin{bmatrix} Q_{8} \\ 0\\ P_{32}\\0 \end{bmatrix},  
\begin{bmatrix} P_{32}  \\ 0\\ Q_{8} \\0 \end{bmatrix},
\begin{bmatrix} 0\\Q_{8}\\0\\P_{32} \end{bmatrix},
\begin{bmatrix} 0 \\P_{32}\\0\\Q_{8} \end{bmatrix} 
 \end{smallmatrix}  \right\} \label{eq_t2_L5}
\end{align}
and 
\begin{align}
L_6 = \left\{  \begin{smallmatrix} 
\begin{bmatrix} R_{8} \\ S_{32}\\0\\0 \end{bmatrix},  
\begin{bmatrix} S_{32}  \\ R_{8} \\ 0 \\0 \end{bmatrix},
\begin{bmatrix} 0\\0 \\R_{8}\\S_{32} \end{bmatrix},
\begin{bmatrix} 0 \\0\\S_{32}\\R_{8} \end{bmatrix} 
\\
\\  
\begin{bmatrix} R_{8} \\ 0\\ S_{32}\\0 \end{bmatrix},  
\begin{bmatrix} S_{32}  \\ 0\\ R_{8} \\0 \end{bmatrix},
\begin{bmatrix} 0\\R_{8}\\0\\S_{32} \end{bmatrix},
\begin{bmatrix} 0 \\S_{32}\\0\\R_{8} \end{bmatrix} 
 \end{smallmatrix}  \right\}. \label{eq_t2_L6} 
\end{align}

%
%
\begin{figure}[!t]
\centering
\includegraphics[width=2.5in,height=2in]{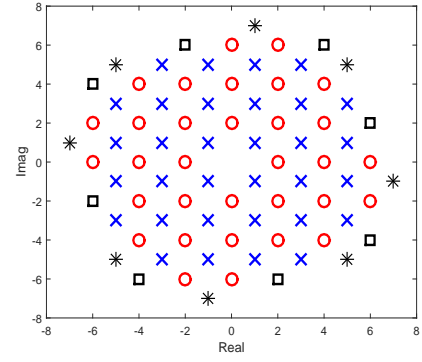}
\caption{The constellations used in ESM-Type2 with $M = 64$. The blue crosses represent 32QAM, the red circles represent $S_{32}$, the black stars represent $R_8$, and the black squares represent $Q_8$.}
\label{F.esm-ty2-64qam}
\end{figure}

The $Q_8$ and $R_8$ constellations are shown in Fig. \ref{F.esm-ty2-64qam} together with $P_{32}$ and $S_{32}$. Mathematically, they can be represented as:
\begin{align}
Q_8 & = \left\{  
	        \begin{matrix} 
			 4+6i,-4-6i,6-4i,-6+4i\\
             6-4i,-6+4i
             6+2i,-6-2i
             \end{matrix} 
             \right 
             \} \notag
\end{align}
and
\begin{align}
R_8 & = \left\{  
	        \begin{matrix} 
			 5+5i,5-5i,
             -5+5i,-5-5i \\
             1+7i,7-1i,
             -1-7i,-7+1i
             \end{matrix} 
             \right 
             \}. \notag
\end{align}

For $M = 16$, the average energy of the transmitted codewords is $12$ in subspaces $L_1$, $L_2$, and $L_3$, because both of the two constellations used in these subspaces have an average energy of $6$. In contrast, the average codeword energy is $16$ in $L_4$, since constellation $Q_4$ has an average energy of $10$. Therefore, the average energy per codeword is given by:
\begin{equation}
E_{ESM-Type2-16QAM} = \frac{3}{4} \times 12 + \frac{1}{4} \times 16 = 13. \notag
\end{equation}
This scheme provides an energy saving of approximately 35\% (13 instead of 20) compared to baseline MSM, which corresponds to a 1.9 dB gain in the decibel scale.

For $M = 64$, symbol selection in signal subspaces $L_1$, $L_2$, and $L_3$ requires $10$ bits. Together with the $2$ prefix bits of the $L_i$ subspaces and the $2$ bits needed for selection of an antenna and constellation combination in the selected subspace, the total number of bits is $14$. In subspace $L_4$, symbol selection requires only $8$ bits, but one additional bit is needed to select $L_5$ or $L_6$, and $3$ bits are needed to select one antenna and modulation combination in the selected $L_i$ subspace. Here too, together with the $2$ prefix bits of the $L_4$ subspace, the number of bits is $14$, and clearly the proposed design achieves $14$ bpcu.

To compute the total energy per transmitted codeword, we first evaluate the average energy of the constellations used in this design: A simple inspection of Figs. \ref{F.esm-type2} and \ref{F.esm-ty2-64qam} shows that the average energy is $E_{P_{32}}=20$ for $P_{32}$, $E_{S_{32}}=22$ for $S_{32}$, $ E_{Q_8}=46$ for $Q_8$, and $E_{R_8}=50$ for $R_8$. Since the symbols take their values from the set ${P_{32},S_{32}}$ in 3 out of the 4 subspaces, from ${P_{32},Q_{8}}$ in one subspace, and from ${S_{32},R_8}$ in the remaining subspace, the average energy per codeword is given by:
\begin{equation}
E_{ESM-Type2-64QAM} = \frac{3}{4}\times 42 + \frac{1}{8}\times 66 + \frac{1}{8} \times 72 = 48.75. \notag
\end{equation}
Compared to the baseline MSM scheme, this ESM design achieves a transmit energy saving of approximately 42\%. This represents an SNR gain of 2.4 dB.

%
%
\begin{figure}[!t]
\centering
\includegraphics[width=2.5in,height=2in]{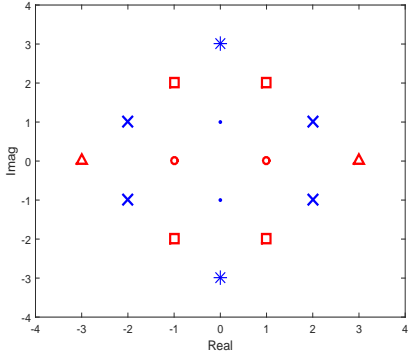}
\caption{The constellations used in ESM-Type3 with $M = 16$. The blue points represent $T_R$, the blue crosses represent $T'_4$, the blue stars represent $T'_2$, the red circles represent $F_R$, the red squares represent $F'_4$, and the red triangles denote $F'_2$.}
\label{F.ESM-type3}
\end{figure}

%
%
\subsection{ESM-Type3}
The reduced-size secondary signal constellation used in our first two designs (ESM-Type1) and ESM-Type2) was derived through a single-step geometric interpolation in the primary constellation plane. Our third design goes one step further and uses two additional signal constellations $T_{M/2}$ and $F_{M/2}$, which are derived through a second interpolation step.

In partitioned form, these constellations are defined as 
\begin{equation}
T_{M/2} = T'_{M/4} \cup T''_{M/8}\cup T_R \notag
\end{equation}
and
\begin{equation}
F_{M/2} = F'_{M/4} \cup F''_{M/8}\cup F_R, \notag
\end{equation}
where $ T'_{M/4}$ denotes the $M/4$ points of $T_{M/2}$ of smallest energy excluding the innermost points, which form $T_c = \{i,-i\}$, $T''_{M/8}$ is the $M/8$ point extension of $T'_{M/4}$ of minimum energy, and $T_R$ denotes the rest of the points in $T_{M/2}$. For the $F_{M/2}$ constellation, we use the same definition and similar notations, because as it will be clear later this constellation is obtained by a simple $\pi/2$ rotation of constellation $T_{M/2}$. The innermost points of $F_{M/2}$ are given by $T_c=\{1,-1\}$. 

For $M=16$, Fig. \ref{F.ESM-type3} shows the 6 component constellations which form $T_{M/2}$ and $F_{M/2}$. They have the following mathematical representation: 
\begin{align*}
T'_4 = \{ \pm 2 \pm i\}, T''_2 = \{\pm 3i\}, T_R = \{ \pm i\}, \, \\
F'_4 = \{ \pm 1 \pm 2i\}, F''_2 = \{\pm 3\}, F_R = \{ \pm 1\}.
\end{align*}

%
%
\begin{figure}[!t]
\centering
\includegraphics[width=2.5in,height=2in]{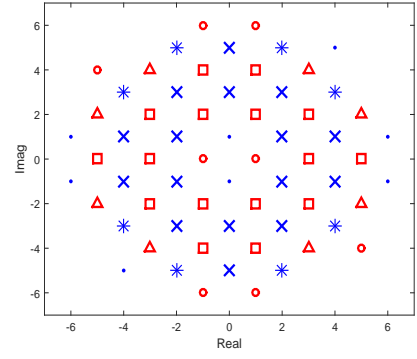}
\caption{The constellations of ESM-Type3 with $M = 64$: The blue points represent $T_R$, the blue crosses represent $T'_{16}$, the blue stars represent $T''_8$, the red circles represent $F_R$, the red squares represent $F'_{16}$, and the red triangles denote $F''_8$.}
\label{F.esm-ty3-64qam}
\end{figure}

For $M=64$, the $6$ component constellations of $T_{32}$ and $F_{32}$ are shown in Fig. \ref{F.esm-ty3-64qam}. Their mathematical representation is:
\begin{align*}
T'_{16} & = \{ \pm 2, \pm 2 \pm i,   \pm 3i, \pm 4 \pm i, \pm 5i \}, \\
T''_8 & = \{ \pm 2 \pm 5i, \pm 4 \pm 3i\}, \\
T_R & = \{ \pm i, \pm 6 \pm i, 4+5i, -4-5i \},
\end{align*}
and 
\begin{align*}
F'_{16} & = \{ \pm 1 \pm 2i, \pm 3 \pm 2i, \pm 1 \pm 4i, \pm 3, \pm 5 \}, \\
F''_{8} &= \{ \pm 3 \pm 4i ,  \pm 5 \pm 2i\}, \\
F_R &= \{ \pm 1, \pm 1 \pm 6i, -5+4i, 5-4i\}. 
\end{align*}
Note that all constellations used in the ESM-Type3 design, i.e. ($P_{M/2},S_{M/2},T_{M/2},F_{M/2}$), have the minimum Euclidean distance of $\delta_0$. Next, the minimum distance between the $P_{M/2}$ and $S_{M/2}$ constellations (resp. the $T_{M/2}$) and $F_{M/2}$ constellations) is $\delta_0/\sqrt{2}$. Finally, the minimum distance between a point taken from $P_{M/2} \cup S_{M/2}$ and a point taken from $T_{M/2}\cup F_{M/2}$ is $\delta_0/2$. Since the number of active antennas is limited to $2$, a particular care must be exercised to preserve a minimum distance of $\delta_0$ in the signal space.

More specifically, the use of different constellations cannot be made independently from a channel use to the next. Instead, the antenna/constellation combinations must be jointly defined over a block of two consecutive channel uses. The minimum distance can be preserved in the following two cases: In the first case, the $P_{M/2}$ and $S_{M/2}$ constellations (resp. the $T_{M/2}$ and $F_{M/2}$ constellations) are employed during both channel uses. In the second case, the $P_{M/2}$ and $S_{M/2}$ constellations are used during the first channel use, and the $T_{M/2}$ and $F_{M/2}$ constellations are used during the second channel use, or vice versa. In this paper, we take the second approach, because the number of bits transmitted per block is not constant in the first.

For presenting our ESM-Type3 scheme, we first extend the system model by stacking two consecutive received signal vectors. Assuming slow-fading channels essentially constant over two consecutive channel uses, the transmitted and received signals are related by the following equation:
\begin{align}
\textbf{Y} = \textbf{H}\textbf{X}+\textbf{N}, 
\end{align}
where $\textbf{Y}=[\textbf{y}_1, \textbf{y}_2]$ denotes the $N_R \times 2$ received signal matrix, $\textbf{X}=[\textbf{x}_1,\textbf{x}_2]$ is the $N_T \times 2$ transmitted signal matrix, $\textbf{N}$ is the $N_R \times 2$ AWGN matrix, and the subscript $k\in \{1,2\}$ denotes the time index of the symbol vector. The transmitted codeword (symbol matrix) $\textbf{X}$ belongs to the following signal space:
%
%
\begin{align}
\mathbf{X}\in \{ \mathbb{S}_{1}, \mathbb{S}_{2}\}, \label{eq.esmty3} 
\end{align}
where
%
%
\begin{align}
& \mathbb{S}_{1} = \{ \textbf{x}_1 \in \mathbb{S}_{PS}, \textbf{x}_2 \in \mathbb{S}_{TF} \}  \\
& \mathbb{S}_{2} = \{ \textbf{x}_1 \in \mathbb{S}_{TF}, \textbf{x}_2 \in \mathbb{S}_{PS} \}.  
\end{align}
In this representation, $\mathbb{S}_{PS}$ denotes the set of symbol vectors
based on the primary and the secondary constellations, and $\mathbb{S}_{TF}$
denotes the set of symbol vectors based on the third and the fourth constellations. The transmitted $N_T \times 2$ codeword takes its values from the set $\mathbb{S}_{PS}$ during the first channel use in the
block and from the set $\mathbb{S}_{TF}$ during the second channel use, or vice versa. The number of bits per codeword is $4 + 4\log_2 M$ , which is twice the number of codewords per channel use. From those, 1 bit selects subset $\mathbb{S}_1$ or subset $\mathbb{S}_2$. Next, $2+2\log_2 M$ bits select a vector from $\mathbb{S}_{PS}$ and $1+2\log_2 M$ bits select a vector from $\mathbb{S}_{TF}$, and these two vectors are transmitted in the order determined by the first bit. 

The details of the proposed design process can be described as follows: First, the set of symbol vectors $\mathbb{S}_{PS}$ is actually the signal space of ESM-Type2 described in the previous subsection. A signal vector from this set is of the form:
\begin{equation}
\mathbb{S}_{PS}: \, \textbf{x} \in \{ L_1,L_2,L_3,L_4 \}, 
\end{equation}
where the subsets $L_1-L_4$ are given by eqns. (\ref{ty2_L1}) $-$ (\ref{eq_esm_ty2_L4}). As shown in the previous subsection, this scheme transmits $2+2\log_2 M$ bits per channel use, and the average total energy per transmitted symbol vector is $E_s = 13$ for $M=16$ and $E_s = 48.75$ for $M = 64$. 

Next, the set of symbol vectors $\mathbb{S}_{TF}$ is based on the third and the fourth constellations $T_{M/2}$ and $F_{M/2}$, but symbol vectors in $\mathbb{S}_{TF}$ transmit one bit less than the $2+2\log_2 M$ bpcu transmitted in the case of $\mathbb{S}_{PS}$. The set $\mathbb{S}_{TF}$ is constructed by the union of four subsets $L'_1$, $L'_2$, $L'_3$, $L'_4$:
\begin{equation}
\mathbb{S}_{TF}: \, \textbf{x} \in \{ L'_1,L'_2,L'_3,L'_4 \}. 
\end{equation}

The first subset is defined as:
\begin{align}
 L'_1 & = \left\{ \begin{smallmatrix}
\begin{bmatrix} T_{M/2} \\ F_{M/2}\\0\\0 \end{bmatrix},  
\begin{bmatrix} F_{M/2}  \\ T_{M/2} \\ 0 \\0 \end{bmatrix},
\begin{bmatrix} 0\\0 \\T_{M/2}\\F_{M/2} \end{bmatrix},
\begin{bmatrix} 0 \\0\\F_{M/2}\\T_{M/2} \end{bmatrix}  \end{smallmatrix} \right\}.   
\end{align}
The $L'_1$ subset can transmit $2+2\log_2 (M/2)$ bits: $2$ bits select one of the four combinations of active TX antennas and associated constellations, $\log_2 (M/2)$ bits select a symbol from the $T_{M/2}$ constellation, and $\log_2 (M/2)$ bits select a symbol from the $F_{M/2}$ constellation. 

The other three subsets $L'_2-L'_4$ are defined as follows:
\begin{align}
L'_2 & = \left\{  \begin{smallmatrix} 
\begin{bmatrix} T'_{M/4} \\ 0 \\ F'_{M/4}\\0 \end{bmatrix},  
\begin{bmatrix} F'_{M/4} \\ 0 \\ T'_{M/4}\\0 \end{bmatrix},
\begin{bmatrix} 0\\T'_{M/4} \\ 0 \\ F'_{M/4} \end{bmatrix},
\begin{bmatrix} 0\\F'_{M/4} \\ 0 \\ T'_{M/4} \end{bmatrix} 
\\
\\  
\begin{bmatrix} T'_{M/4} \\ 0 \\ 0 \\ F'_{M/4} \end{bmatrix},  
\begin{bmatrix} F'_{M/4} \\ 0 \\ 0 \\ T'_{M/4}  \end{bmatrix},
\begin{bmatrix} 0 \\ T'_{M/4} \\ F'_{M/4} \\ 0  \end{bmatrix},
\begin{bmatrix} 0 \\ F'_{M/4} \\ T'_{M/4} \\ 0  \end{bmatrix} 
 \end{smallmatrix}  \right\} \\
L'_3 &= \{ L'_2 | T'_{M/4} \rightarrow T''_{M/8}\} \\ 
L'_4 &= \{ L'_2 | F'_{M/4} \rightarrow F''_{M/8}\}.
\end{align}
This representation indicates that the combinations of active antennas in subset $L'_3$ are the same as those in $L'_2$, but here constellation $T'_{M/4}$ is replaced by constellation $T''_{M/8}$. Similarly, subset $L'_4$ is obtained from $L'_2$ by substituting constellation $F''_{M/8}$ for $F'_{M/4}$. 

The signal subset $L'_2$ transmits $3+2\log_2 (M/4)$ bits per symbol vector: 3 bits are needed to select one of the 8 combinations, $\log_2 (M/4)$ bits to select a symbol from $T'_{M/4}$, and $\log_2 (M/4)$ bits to select a symbol from $F'_{M/4}$. Next, since the $L'_3$ subset is derived from $L'_2$ by substituting $T''_{M/8}$  for $T'_{M/4}$, it transmits $3+\log_2 (M/4)+\log_2 (M/8)$ bits per symbol vector. Again, 3 bits select one of the 8 combinations, and then $\log_2 (M/8)$ bits select a symbol from $T''_{M/8}$, and $\log_2 (M/4)$ bits select a symbol from $F'_{M/4}$. Similarly, $L'_4$ subset transmits $3+\log_2 (M/4)+ \log_2 (M/8)$ bits per symbol vector. Here, 3 bits select one of the 8 combinations, $\log_2 (M/8)$ bits select a symbol from $F''_{M/8}$, and $\log_2 (M/4)$ bits select a symbol from $T'_{M/4}$.

The discussion above indicates that the number of bits transmitted per symbol vector is not uniform across the $L'_1-L'_4$ subsets. The implication of this is that the prefix of these subsets in $\mathbb{S}_{TF}$ must have a variable number of bits. Specifically, subset $L'_1$ must have a 1-bit prefix, subset $L'_2$ a 2-bit prefix, and subsets $L'_3$ and $L'_4$  must have a 3-bit prefix. With these variable-length prefixes, it can be seen that all symbol vectors in $\mathbb{S}_{TF}$ carry $1+2\log_2 M$ bits. 

At this point, it is important to clarify the difference between the construction of the $L'_1$ subset and that of the $L'_2- L'_4$ subsets included in $\mathbb{S}_{TF}$. Notice that the innermost points of the $T_{M/2}$ and $F_{M/2}$ constellations, namely $T_c$ and $F_c$, are only used in the first subset $L'_1$. These points cannot be used in $L'_2$, because otherwise the minimum Euclidean distance in the signal space would be $\delta_0/\sqrt{2}$, which is 3 dB smaller than the minimum Euclidean distance in $\mathbb{S}_{PS}$. This is the case, for instance, between the symbol vectors $[1,i,0,0] \in L'_1$ and $[1,0,i,0] \in L'_2$. Similarly, the innermost points are not allowed in subsets $L'_3$ and $L'_4$. As a result, the signal vectors in $\mathbb{S}_{TF}$ carry only $1+2\log_2 M$ bits, while the signal vectors in $\mathbb{S}_{PS}$ carry $2+2\log_2 M$ bits.

For $M=16$, the average energy per transmitted symbol vector from $\mathbb{S}_{TF}$ is $E_s=11$. Since the signal vector sets in $\mathbb{S}_{PS}$ and $\mathbb{S}_{TF}$ are used with the same probability, the average energy of the transmitted codewords in ESM-Type3 is: 
\begin{equation}
E_{ESM-Type3-16QAM} = \frac{1}{2}\times (13+11) = 12. \notag
\end{equation}
This represents a 2.2 dB SNR gain over MSM and a 0.4 dB gain over the ESM-Type2.

For $M=64$, the average energy per transmitted symbol vector from $\mathbb{S}_{TF}$ is $E_s= 37$, and the average energy of the transmitted ESM-Type3 codewords is:
\begin{equation}
E_{ESM-Type3-64QAM} = \frac{1}{2}\times (48.75+37) = 42.875. \notag
\end{equation}
This represents a 2.9 dB SNR gain over baseline MSM.

%
%
\section{Extensions to Higher Number of Antennas}
In this section, we investigate the extension to higher numbers of antennas of the new ESM designs presented in the previous section. Before doing this, we describe the MSM concept used for benchmarking these designs. In MSM with $N_T$ transmit antennas out of which $N_A$ antennas are active using $M$-QAM modulation, the maximum number of active antenna combinations is $C_{N_T}^{N_A}=\frac{N_T!}{N_A!\times (N_T-N_A)!}$. Usually, the number of combinations is restricted to be an integer power of 2 in order to have an integer number of address bits to select the active antennas. This number is given by:
\begin{equation}
n = \lfloor \log_2 (C^{N_A}_{N_T})\rfloor,
\end{equation}
where $\lfloor x \rfloor$ stands for the integer part of $x$. In this scheme, the transmitted average energy is $10 \times N_A$ for $M=16$ (16QAM modulation) and $42\times N_A$ for $M=64$ (64QAM modulation). As for the throughput, it is given by $n + N_A\log_2 M$.

\subsection{ESM-Type1}
The basic principle of ESM-Type1 is to use a secondary constellation of half size (with $M/2$ points) in addition to the primary constellation with $M$ points in order to reduce the average transmit energy. The primary constellation is the $M$-QAM constellation (denoted $P_M$) used by the reference MSM scheme, and the secondary constellation is the $S_{M/2}$ constellation. Selection of the active antennas requires the same number of address bits as in MSM. But on top of this, ESM-Type1 requires additional address bits to select the antennas which transmit symbols from the $S_{M/2}$ constellation. 

Assuming that the number of active antennas $N_A$ is an even number, half of the active antennas transmit symbols from the $P_M$ constellation, and the other half of the antennas transmit symbols from the $S_{M/2}$ constellation. Compared to MSM, the number of bits in the transmitted symbols is reduced by $N_A/2$. This reduction is compensated by the bits assigned to the selection of the active antennas which transmit symbols from the $S_{M/2}$ constellation. For a given set of active antennas, the number of bits assigned to this selection is 
\begin{equation}
m = \left \lfloor \log_2 \left( C^{N_A/2}_{N_A} \right) \right \rfloor.
\end{equation}
For example, with $N_A=4$, we have $m=2$, and precisely, this is the number of bits that we need to compensate for the fact that the $4$ symbols in ESM-Type1 transmit $2$ bits less than in MSM. Consequently, in this scheme too the throughput is given by $ n + N_A\log_2 M$.

The average transmit energy is clearly $10 \frac{N_A}{2}+6\frac{N_A}{2} =16\frac{N_A}{2}$ for $M = 16$, and $42\frac{N_A}{2}+22\frac{N_A}{2} =64\frac{N_A}{2}$ for $M = 64$. For all $N_A$ values, the gain with respect to MSM is 1 dB and 1.2 dB with $M=16$ and $M=64$, respectively.

\subsection{ESM-Type2}
The idea here is not to use the original primary constellation $P_M$ in full, but instead a subset $P_{M/2}$, which consists of the $M/2$ points of $P_M$ of smallest energy. With $N_A$ active antennas, the number of bits carried by the transmitted symbols is reduced by $N_A$ with respect to MSM which uses the original $P_M$ constellation. Since both of the constellations used in this design have the same size and essentially the same average energy, we do not need to restrict here that half of the symbols must take their values from $P_{M/2}$ and the other half from the $S_{M/2}$ constellation. All we need instead is to have an even number of symbols taking their values from $S_{M/2}$, as this condition is sufficient to ensure that the minimum Euclidean distance in the signal space will not be reduced. The group of bits assigned to the selection of the constellation must form a parity-check code and hence it contains $N_A-1$ information bits. This compensates for the loss of $N_A$ bits due to the half-size constellations, except for $1$ bit. Compensation of this bit can only be made by increasing the number of active antenna combinations and adding some other combinations which make use of additional modulations, as illustrated by the signal space in section III.C.

We now illustrate the signal space construction for $N_T=8$ and three different values of $N_A$, namely $N_A=2$, $N_A=4$, and $N_A=6$. For both $N_A=2$ and $N_A=6$, the maximum number of active antenna combinations is $C_8^2=28$. From those, MSM uses $16$, which require $4$ address bits. In contrast, ESM-Type2 uses all of these combinations, and in addition to them, it uses additional antenna/modulation combinations which involve other constellations, similar to the subspace given by (\ref{eq_esm_ty2_L4}) for $M = 16$ and to the subspaces given by (\ref{eq_t2_L5}) and (\ref{eq_t2_L6}) for $M = 64$. It can be easily verified that the average energy per codeword is given by $(28\times 12 + 4 \times 16)/32 = 12.5$ for $M = 16$, and $(28\times 42 + 2\times 66 + 2\times 72)/32=45.375$ for $M = 64$. Note that the energy saving with respect to MSM here is higher than that reported in Section III.C. More specifically, the energy saving is 2.04 dB for $M = 16$ and 2.7 dB for $M = 64$. 

For $N_T=8$ and $N_A=4$, the situation is not as favorable: Indeed, the number of active antenna combination is $C_8^4=70$, and MSM can use $64$ of them. Instead of trying to find suitable antenna and modulation combinations to increase the signal space and recover the missing bit, we found that in this case a simple alternative consists of using constellation $S_{M/2}$ on two antennas, constellation $P_{M/2}$ on one antenna, and the full constellation $P_M$ on the remaining active antenna. The energy saving with respect to MSM in this case is 1.55 dB for $M=16$ and 2.0 dB for $M=64$, which is a worst-case situation corresponding to one of the active antennas transmitting symbols from the full primary constellation. In summary, the gain achieved with respect to MSM is a function of the $N_T$ and $N_A$ parameters, and it will exceed 2 dB in most cases.

\subsection{ESM-Type3}
We will not attempt here to fully describe ESM-Type3 for an arbitrary number of transmit antennas $N_T$ and an arbitrary number of active antennas $N_A$, because the signal space will depend on both of these parameters. Instead, we will give the basic design rule and indicate the achievable performance.  

Recall that this ESM design makes use of 4 different constellations, namely $P_{M/2}$, $S_{M/2}$, $T_{M/2}$, and $F_{M/2}$, the first being a subset of the primary constellation, the second being a secondary constellation derived through a first interpolation step, and finally the third and the fourth being derived through a second interpolation step. Also recall that all of these modulations have a minimum Euclidean distance of $\delta_0$, the minimum distance between $P_{M/2}$ and $S_{M/2}$ (resp. between $T_{M/2}$ and $F_{M/2}$ is  $\delta_0/\sqrt{2}$, and the minimum distance between $P_{M/2} \cup S_{M/2}$ and $T_{M/2} \cup F_{M/2}$ is $\delta_0/2$. 

Let us define 2 bit sequences $\{\alpha_i\}$ and $\{ \beta_i \}$, $i=1,2,\cdots,N_A$ where $\alpha_i$ determines whether the $i$th component of the codeword belongs to $P_{M/2} \cup S_{M/2}$ or to $T_{M/2}\cup F_{M/2}$, and $\beta_i$ determines whether this component belongs to $P_{M/2}\cup T_{M/2}$ or to $S_{M/2}\cup F_{M/2}$. In order to preserve a minimum distance of  $\delta_0$ in the signal space, the $\{\alpha_i\}$ sequence must form a binary code of Hamming distance 4, and the $\{\beta_i\}$ sequence must form a binary code of Hamming distance 2. With $N_A = 2$, a Hamming distance of 4 cannot be achieved if the codewords are defined over a single channel use, and for this reason two consecutive symbol vectors were stacked and the codewords were defined over two consecutive channel uses in Section III.C. This constraint remains with higher $N_T$ values as long as $N_A=2$. But for $N_A$ values of 4 or higher, no stacking is required, because a Hamming distance of 4 can be achieved between $\{\alpha_i \}$, $i=1,2,\cdots,N_A$ sequences defined over a single channel use. The design rule in ESM-Type3 is to define the signal space in such a way that these two Hamming distance requirements are met. Then, the SNR gain over MSM is obtained simply by comparing the average transmit energies.

\section{Performance and Complexity Analysis}
\subsection{The Minimum Euclidean Distance}
Assuming the channel state information (CSI) is perfectly known at the receive side, the maximum-likelihood (ML) decoder estimates the transmitted codeword according to:
%
%
\begin{align}
\mathbf{\hat X}  = \arg \min_{\mathbf{X} \in \mathbb{X}} \Vert \mathbf{Y}-\mathbf{H}\mathbf{X}\Vert^2,
\end{align}
where the minimization is performed over all possible codewords from the signal space $\mathbb{X}$. 

In ML detection using exhaustive search, the receiver computes the Euclidean distance between the received noisy signal and the set of all possible codewords transmitted over the channel matrix. At high SNR, the receiver performance is dominated by the minimum squared Euclidean distance over the signal space \cite{stbc_05}:
%
%
\begin{align}
L^2_{min} = \min \Vert \textbf{X}-\textbf{X}' \Vert^2.
\end{align}
The ESM schemes introduced in this paper were designed in such a way as to preserve the minimum squared Euclidean distance $\delta_0$ of the primary modulation, i.e., $L_{min}^2 = \delta_0^2$ in all of them. The same minimum distance being also valid for single-stream SM, MSM, the ESM schemes introduced in \cite{esm_15}, and in SMX, comparison of the respective asymptotic performances of the different schemes is reduced to comparing their average transmit energy $E_s$. The average transmit energy for all of these MIMO schemes is summarized in Table \ref{t.eg} for 10 bpcu and for 14 bpcu transmissions. 
\begin{table}
\renewcommand{\arraystretch}{1.3}
\caption{Average transmit energy for 10 bpcu and 14 bpcu}
\centering
\begin{tabular}{|c||c|c|c|c|}
\hline 
 & SM \cite{spa_08v} & SMX-2TX & ESM \cite{esm_15} & MSM \cite{per_13}\\ 
\hline \hline 
10bpcu & 170 & 40 & 28.5 & 20 \\ 
\hline 
14bpcu & 2730 & 164 & 202 & 84\\ 
\hline
\hline 
 & \textbf{ESM-Type1} & \textbf{ESM-Type2} & \textbf{ESM-Type3} & SMX-4TX\\
 \hline
10bpcu & 16 & 13 & 12 & 16 \\ 
\hline
14bpcu & 64 & 48.75 & 42.875 & 32\\
\hline
\end{tabular} 
\label{t.eg}
\end{table} 

The gains achieved by the new ESM designs over MSM have already been indicated in Section III. The main purpose of this table is to give an indication as to how these schemes compare to spatial multiplexing with 4 transmit antennas (SMX-4TX), spatial multiplexing with 2 transmit antennas (SMX-2TX), single-stream SM of \cite{spa_08v}, and also to the original ESM schemes of \cite{esm_15}. First, note that SMX-4TX must use two different modulations at these two spectral efficiencies. For 10 bpcu transmission, we assume that 2 antennas transmit QPSK symbols (of average energy 2) and the other 2 antennas transmit symbols from the $P_8$ constellation used by ESM-Type2 (see Subsection III.C). The average transmit energy of SMX-4TX is $2\times(2+6)=16$ in this case. For 14 bpcu transmission, 2 antennas transmit 16QAM symbols and the other 2 antennas transmit symbols from the $P_8$ constellation. The average transmit energy is $2\times(10+6)=32$. Clearly, this transmission scheme has better performance than our new ESM schemes at 14 bpcu, but it involves 4 RF chains

Next, SMX-2TX uses 32QAM for 10 bpcu and 128QAM for 14 bpcu transmission. The average transmit energy is 40 and 164, respectively. The gains achieved by our new ESM designs over this scheme are substantial: around $4.0 - 4.1$ dB with ESM-Type1, $4.9 - 5.3$ dB with ESM-Type2, and $5.2 - 5.8$ dB with ESM-Type3.

Single-stream SM must employ 256QAM modulation to achieve 10 bpcu and 4096QAM to achieve 14 bpcu. It is needless to say that the gap is tremendous here. Finally, our original ESM scheme of \cite{esm_15} achieves 10 bpcu using 64QAM as primary modulation and 14 bpcu using 1024QAM as primary modulation. In the first case, it uses two secondary modulations of 8 points, and in the second case, it uses secondary modulations of 32 points each, following the design rules described in that paper. The average transmit energy values given in Table \ref{t.eg} indicate that in the case of 10 bpcu transmission ESM-Type3 gains $10 \log_{10}(28.5/12)=3.8$ dB over our original ESM scheme. In the case of 14 bpcu, the gain is as high as 6.7 dB. These results are not surprising, because the ESM scheme of \cite{esm_15} was designed to improve over single-stream SM, while the new ESM schemes introduced in this paper were specifically designed to improve over MSM. 

\subsection{The Union Bound Analysis}
For each channel use, the signal codeword $\mathbf{X}$ is in a vector form $\mathbf{x}$ and its performance can be evaluated by using the union bound analysis shown in \cite{esm_15}. We define the pairwise error probability (PEP) as the probability that the ML decoder decodes a symbol vector $\mathbf{x}′$ instead of the transmitted symbol vector $\mathbf{x}$. The average PEP (APEP) can be computed by using the union bound as follows:
%
%
\begin{align}
APEP\leq \frac{1}{\vert \mathbb{X} \vert} \sum_{\textbf{x}\in \mathbb{X}}  \sum_{\textbf{x}'\in \mathbb{X}} PEP(\textbf{x}\rightarrow \textbf{x}'). \label{eq_apep}
\end{align}
For Rayleigh fading channels, the PEP is given by
%
%
\begin{align}
& PEP(\textbf{x}\rightarrow \textbf{x}')\notag \\ & = \mathbb{E}_{\textbf{H}}\left [ \mathcal{Q}\left( \sqrt{\frac{E_s \Vert \textbf{H}\textbf{x}-\textbf{H}\textbf{x}' \Vert^2}{2N_0}} \right) \right] \notag\\ 
&  = \left( \frac{1-\mu}{2}\right)^{N_R} \sum^{N_R-1}_{k=0}C ^k_{N_R-1+k}\left(\frac{1+\mu}{2}\right)^k,
\end{align}
where the Gaussian Q-function is denoted by $\mathcal{Q}(\cdot)$, $\mu = \sqrt{\tau/(4N_0/E_s + \tau)}$, and $\tau = \Vert \textbf{x} - \textbf{x}' \Vert^2$ denotes the squares Euclidean distance between two symbol vectors.

The APEP shown in (\ref{eq_apep}) can be used for an analytic evaluation of the proposed ESM schemes. Given the codeword length with $N_c$ channel uses, the codeword error rate (CER) can be upper bounded by:
\begin{equation}
CER \leq N_c \times APEP. \label{eq_cer}
\end{equation}
For ESM-Type1 and ESM-type2, error events are independent from a channel use to the next, because each symbol vector is generated independently. Therefore, the CER is bounded by the product of the APEP per channel use and the codeword length $N_c$. For ESM-Type3, a codeword is composed of two symbol vectors transmitted over two channel uses. These two symbol vectors have the same error rate, due to the symmetry imposed on the signal design. As a result, the CER of ESM-Type3 can also be bounded using equation (\ref{eq_cer}).   

\subsection{Receiver Complexity}
We define the receiver complexity as the number of floating point operations (flops) required per ML decoder decision, where each addition, subtraction, multiplication, division, and square-root operation counts as one flop \cite{flop_13}. Using this definition, we found that the first two of the proposed ESM schemes have essentially the same receiver complexity as MSM, while the third has a 50\% higher complexity. 

Using the system model given by eqn. (\ref{eq.system_model}), the ML decoder needs to compute $2^b$ decision metrics $w_k = \Vert \textbf{y} - \textbf{H}\textbf{x}_k \Vert^2$, $k=1,\cdots,2^b$, where $b$ is the total number of transmitted bits per channel use. This holds for MSM as well as for ESM-Type1 and ESM-Type2. For different operations, the number of flops is given by:
\begin{itemize}
\item 
Computing $\textbf{H}\textbf{x}_k$ requires $N_R(2N_A-1)$ flops,
\item
Computing $\textbf{y}-\textbf{Hx}_k$ requires $N_R$ flops,
\item
Computing $\Vert \textbf{y} - \textbf{Hx}_k \Vert^2$ requires $2N_R-1$ flops.
\end{itemize}
That is, computation of the decision metrics by the ML decoder requires in total  $2^b (2N_R (N_A+1)-1)$ flops. 

A close look at ESM-Type3 reveals that the decoder complexity is more involved than in the first two ESM schemes, because the ML decoder must jointly decide two consecutive symbols. The ML decoder must search in this space using two consecutive received signal samples $\textbf{y}_1$ and $\textbf{y}_2$ and computing metrics of the form $w_k = \Vert \textbf{y}_1 - \textbf{H}\textbf{x}_i \Vert^2 + \Vert \textbf{y}_2 - \textbf{H}\textbf{x}_j \Vert^2$, where $\textbf{x}_i \in \mathbb{S}_{PS}$, $\textbf{x}_j \in \mathbb{S}_{TF}$, or $\textbf{x}_i \in \mathbb{S}_{TF}$, $\textbf{x}_j \in \mathbb{S}_{PS}$. The number of flops per decoder decision is $2(2^b + 2^{b-1})(2N_R(N_A+1)-1)$. But since only one decision is made every two channel uses, the number of flops per channel use is $(2^b + 2^{b-1})(2N_R(N_A+1)-1)$. This is 50\% higher than in MSM, ESM-Type1, and ESM-Type2.

\subsection{Sphere Decoding for ESM}
Implementation of the ML decoder using exhaustive search involves a very high complexity and becomes prohibitive at very high spectral efficiencies, and this holds for any MIMO scheme. In practice, the ML decoder can be implemented efficiently using the sphere decoding (SD) technique. This technique reduces the complexity of the ML decoder by shrinking the search space to an acceptable level and counting those combinations that lie within a sphere centered on the received signal. The general SD scheme for SM was described in \cite{gen_13s}, where it was shown that this decoding technique significantly reduces the computational complexity with no performance loss. In the simulations section which follows, we use a multi-stream complex-valued SD for ESM, which is a modification of the single-stream and real-valued SD \cite{gen_13s} that takes the signal space of ESM into account and uses an infinite search radius to guarantee the ML performance. 

\section{Simulation Results}
Monte Carlo simulations were carried out using uncorrelated Rayleigh fading MIMO channels and assuming perfect CSI at the receiver. In the simulations, symbol codewords $\textbf{X}$ were randomly generated transmitted over the channel, the SD was performed using the received noisy signal samples, and error events $\textbf{X} \neq \textbf{X}'$ were counted. The obtained codeword error rate (CER) was used to compare baseline MSM and the presented ESM schemes.  

%
%
\begin{figure}[!t]
\centering
\includegraphics[width=3.5in,height=2.5in]{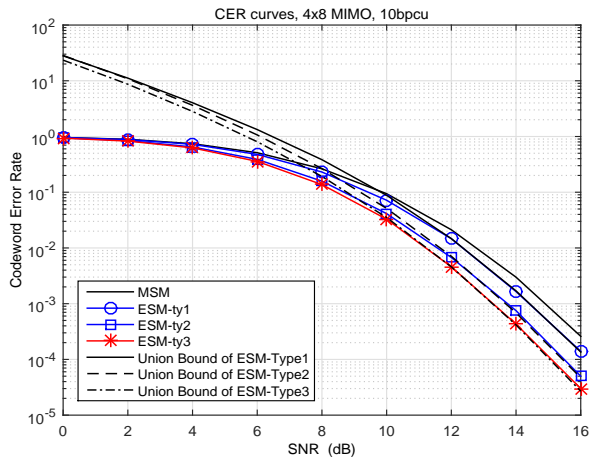}
\caption{The CER performance of MSM and of the proposed ESM schemes: 4 TX antennas and 8 RX antennas with 10 \textit{bpcu}.}
\label{F.pep_10b}
\end{figure}

Fig. \ref{F.pep_10b} gives the Monte-Carlo simulation results of the system performance for 10-bpcu transmission. The number of receive antennas used in these simulations is 8. These results show that at $CER = 10^{-3}$ the presented ESM schemes achieve SNR gains over MSM of around 0.6 dB, 1.3 dB, and 1.8 dB, respectively. In this figure, we also give the analytic bound of the ESM schemes obtained given by (\ref{eq_cer}) to show its tightness in the high SNR region.

%
%
\begin{figure}[!t]
\centering
\includegraphics[width=3.5in,height=2.5in]{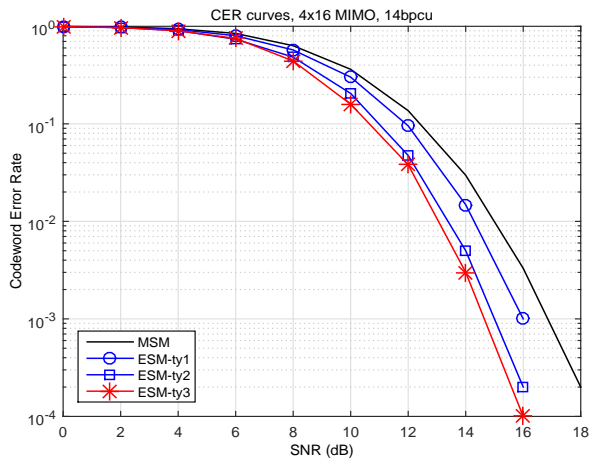}
\caption{The CER performance of MSM and of the proposed ESM schemes: 4 TX antennas and 16 RX antennas with 14 \textit{bpcu}.}
\label{F.pep_11b}
\end{figure}

In Fig. \ref{F.pep_11b}, we report the CER performance of MSM and the proposed ESM schemes providing 14 bpcu using 16 receive antennas. Here, we can see that at the $CER = 10^{-3}$, the ESM schemes achieve gains of around 0.9 dB, 1.9 dB, and 2.2 dB, respectively, over MSM. Note that the gains are higher than those achieved in the 10 bpcu case. This is due to the fact that the average energy of the secondary constellations used in our signal design becomes lower (relatively to the primary constellation) when higher spectral efficiencies are considered. Also note that the gains observed in these simulations are lower than those predicted by the average transmit powers, but this is not surprising, because the latter are asymptotic results that are valid at high SNR values.

%
%
\begin{figure}[!t]
\centering
\includegraphics[width=3.5in,height=2.5in]{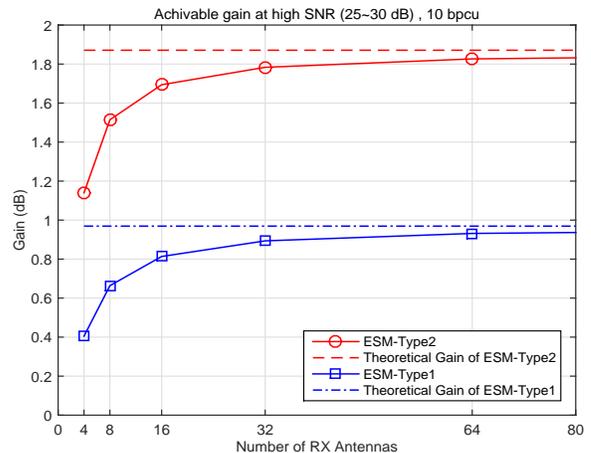}
\caption{Performance gain vs. the number of RX antennas for ESM-Type1 and ESM-Type2 with 10 \textit{bpcu}.}
\label{F.ant_imp}
\end{figure}

A final investigation in this work concerned the evaluation of the number of RX antennas required to approach the gains predicted by the average transmit energies. The results corresponding to ESM-Type1 and ESM-Type2 with 10 bpcu are reported in Fig. \ref{F.ant_imp}. The specific numbers of RX antennas used in this investigation were 4, 8, 16, 32, and 64. The results show that a large number of RX antennas are needed in order to closely approach the asymptotic performance gain, but 80\% of this gain in ESM-Type1 and 90\% in ESM-Type2 can be achieved with 16 RX antennas.

This can be interpreted by using the union bound equation shown in (\ref{eq_apep}). As the number of receiving antennas grows large with a high SNR value, the union bound on the error probability depends only on the average transmit energy and the minimum Euclidean distance between all pairs of codewords, i.e., $PEP_w = \mathcal{Q}\left(\frac{E_s \cdot \delta_0}{\sqrt{2N_0 }}\right)$ when $N_R \rightarrow \infty$ and $N_T$ is finite \cite{large_no_02}. Since the the minimum distance $\delta_0$ is the same for all schemes in this paper, this result shows that the theoretical gain can be achieved with a large number of RX antennas and a high SNR value.

%
%
\section{Conclusion}
Taking multi-stream SM as reference, we introduced in this paper three new ESM designs which lead to increasing SNR gains. The new schemes were described for MIMO systems with 4 transmit antennas two of which remain systematically active, but their extension to higher number of antennas was also presented. The proposed designs extend our previous work reported in \cite{esm_15}, and are based on the concept of multiple constellations. The basic principle is to increase the signal space using additional signal constellations to the primary constellation used by MSM. The first and the second ESM schemes make use of a secondary constellation, which is obtained through a single step of geometric interpolation in the primary constellation plane. The third ESM scheme goes one step further and uses two additional constellations derived through a second interpolation step. In all of them, the signal space is designed in such a way as to preserve the minimum Euclidean distance of the primary constellation while reducing the average total transmit energy. This makes performance comparisons with MSM and other MIMO schemes such as spatial multiplexing straightforward. Focusing on spectral efficiencies of 10 bpcu and 14 bpcu and using Monte Carlo simulations on Rayleigh fading channels as well as analytic performance bounds, it was found that the proposed schemes achieve significant performance gains compared to MSM with two active TX antennas.

%
%
\section*{Acknowledgment}
The present work was carried out within the framework of Celtic-Plus SHARING project.

%
%
\bibliographystyle{IEEEtran}
\bibliography{IEEEabrv,ref}

\begin{thebibliography}{10}
\providecommand{\url}[1]{#1}
\csname url@samestyle\endcsname
\providecommand{\newblock}{\relax}
\providecommand{\bibinfo}[2]{#2}
\providecommand{\BIBentrySTDinterwordspacing}{\spaceskip=0pt\relax}
\providecommand{\BIBentryALTinterwordstretchfactor}{4}
\providecommand{\BIBentryALTinterwordspacing}{\spaceskip=\fontdimen2\font plus
\BIBentryALTinterwordstretchfactor\fontdimen3\font minus
  \fontdimen4\font\relax}
\providecommand{\BIBforeignlanguage}[2]{{%
\expandafter\ifx\csname l@#1\endcsname\relax
\typeout{** WARNING: IEEEtran.bst: No hyphenation pattern has been}%
\typeout{** loaded for the language `#1'. Using the pattern for}%
\typeout{** the default language instead.}%
\else
\language=\csname l@#1\endcsname
\fi
#2}}
\providecommand{\BIBdecl}{\relax}
\BIBdecl

\bibitem{fun_05}
D.~Tse and P.~Viswanath, \emph{Fundamentals of Wireless Communications}.\hskip
  1em plus 0.5em minus 0.4em\relax Cambridge University Press, 2005.

\bibitem{wc_05}
A.~Goldsmith, \emph{Wireless Communications}.\hskip 1em plus 0.5em minus
  0.4em\relax Cambridge University Press, 2005.

\bibitem{mul_09}
J.~Mietzner, R.~Schober, L.~Lampe, W.~Gerstacker, and P.~Hoeher,
  ``Multiple-{A}ntenna {T}echniques for {W}ireless {C}ommunications - {A}
  {C}omprehensive {L}iterature {S}urvey,'' \emph{IEEE Communications Surveys \&
  Tutorials}, vol.~11, no.~2, pp. 87--105, 2009.

\bibitem{spa_08v}
R.~Mesleh, H.~Haas, S.~Sinanovic, C.~W. Ahn, and S.~Yun, ``Spatial
  {M}odulation,'' \emph{IEEE Transactions on Vehicular Technology}, vol.~57,
  no.~4, pp. 2228--2241, 2008.

\bibitem{spa_11}
M.~Di~Renzo, H.~Haas, and P.~M. Grant, ``Spatial {M}odulation for
  {M}ultiple-{A}ntenna {W}ireless {S}ystems: {A} {S}urvey,'' \emph{IEEE
  Communications Magazine}, vol.~49, no.~12, pp. 182--191, 2011.

\bibitem{sin_11}
A.~Mohammadi and F.~Ghannouchi, ``Single {RF} {F}ront-{E}nd {MIMO}
  {T}ransceivers,'' \emph{IEEE Communications Magazine}, vol.~49, no.~12, pp.
  104--109, 2011.

\bibitem{spa_01}
Y.~Chau and S.-H. Yu, ``Space {M}odulation on {W}ireless {F}ading {C}hannels,''
  in \emph{Proc. IEEE VTC 2001- Fall}, vol.~3, 2001, pp. 1668--1671 vol.3.

\bibitem{gen_10g}
J.~Fu, C.~Hou, W.~Xiang, L.~Yan, and Y.~Hou, ``Generalized {S}patial
  {M}odulation with {M}ultiple {A}ctive {T}ransmit {A}ntennas,'' in \emph{Proc.
  IEEE Globecom 2010 Workshops}, 2010.

\bibitem{gen_10}
A.~Younis, N.~Serafimovski, R.~Mesleh, and H.~Haas, ``Generalised {S}patial
  {M}odulation,'' in \emph{Proc. Asilomar Conference on Signals, Systems and
  Computers (ASILOMAR'10)}, 2010.

\bibitem{gen_12}
J.~Wang, S.~Jia, and J.~Song, ``Generalised {S}patial {M}odulation {S}ystem
  with {M}ultiple {A}ctive {T}ransmit {A}ntennas and {L}ow {C}omplexity
  {D}etection {S}cheme,'' \emph{IEEE Transactions on Wireless Communications},
  vol.~11, no.~4, pp. 1605--1615, 2012.

\bibitem{spa_14}
M.~Di~Renzo, H.~Haas, A.~Ghrayeb, S.~Sugiura, and L.~Hanzo, ``Spatial
  modulation for generalized {MIMO}: Challenges, opportunities, and
  implementation,'' \emph{Proceedings of the IEEE}, vol. 102, no.~1, pp.
  56--103, 2014.

\bibitem{spa_09}
J.~Jeganathan, A.~Ghrayeb, L.~Szczecinski, and A.~Ceron, ``Space {S}hift
  {K}eying {M}odulation for {MIMO} {C}hannels,'' \emph{IEEE Transactions on
  Wireless Communications}, vol.~8, no.~7, pp. 3692--3703, 2009.

\bibitem{gen_08}
J.~Jeganathan, A.~Ghrayeb, and L.~Szczecinski, ``Generalized {S}pace {S}hift
  {K}eying {M}odulation for {MIMO} {C}hannels,'' in \emph{Proc. {IEEE}
  Personal, Indoor and Mobile Radio Communications (PIMRC'08)}, 2008.

\bibitem{coh_10}
S.~Sugiura, S.~Chen, and L.~Hanzo, ``Coherent and {D}ifferential {S}pace-{T}ime
  {S}hift {K}eying: {A} {D}ispersion {M}atrix {A}pproach,'' \emph{IEEE
  Transactions on Communications}, vol.~58, no.~11, pp. 3219--3230, 2010.

\bibitem{stb_11}
E.~Basar, U.~Aygolu, E.~Panayirci, and H.~Poor, ``{S}pace-{T}ime {B}lock
  {C}oded {S}patial {M}odulation,'' \emph{IEEE Transactions on Communications},
  vol.~59, no.~3, pp. 823--832, March 2011.

\bibitem{gen_11}
S.~Sugiura, S.~Chen, and L.~Hanzo, ``{G}eneralized {S}pace-{T}ime {S}hift
  {K}eying {D}esigned for {F}lexible {D}iversity-, {M}ultiplexing- and
  {C}omplexity-{T}radeoffs,'' \emph{IEEE Transactions on Wireless
  Communications}, vol.~10, no.~4, pp. 1144--1153, 2011.

\bibitem{spa_14c}
M.-T. Le, V.-D. Ngo, H.-A. Mai, X.~N. Tran, and M.~Di~Renzo, ``Spatially
  {M}odulated {O}rthogonal {S}pace-{T}ime {B}lock {C}odes with
  {N}on-{V}anishing {D}eterminants,'' \emph{IEEE Transactions on
  Communications}, vol.~62, no.~1, pp. 85--99, 2014.

\bibitem{per_13}
K.~Ntontin, M.~Di~Renzo, A.~Perez-Neira, and C.~Verikoukis, ``Performance
  {A}nalysis of {M}ultistream {S}patial {M}odulation with
  {M}aximum-{L}ikelihood {D}etection,'' in \emph{Proc. IEEE Global
  Communications Conference (GLOBECOM'13)}, 2013.

\bibitem{gen_13s}
A.~Younis, S.~Sinanovic, M.~Di~Renzo, R.~Mesleh, and H.~Haas, ``{G}eneralised
  {S}phere {D}ecoding for {S}patial {M}odulation,'' \emph{IEEE Trans. on
  Communications}, vol.~61, no.~7, pp. 2805--2815, July 2013.

\bibitem{esm_15}
C.-C. Cheng, H.~Sari, S.~Sezginer, and Y.~T. Su, ``Enhanced {S}patial
  {M}odulation with {M}ultiple {S}ignal {C}onstellations,'' \emph{IEEE
  Transactions on Communications}, vol.~63, no.~6, pp. 2237--2248, June 2015.

\bibitem{esm_15p}
C.~C. Cheng, H.~Sari, S.~Sezginer, and Y.~T. Su, ``New {S}ignal {D}esign for
  {E}nhanced {S}patial {M}odulation with {M}ultiple {S}ignal
  {C}onstellations,'' in \emph{Proc. PIMRC 2015}, Sept 2015.

\bibitem{stbc_05}
H.~Jafarkhani, \emph{Space-{T}ime {C}oding: {T}heory and {P}ractice}.\hskip 1em
  plus 0.5em minus 0.4em\relax Cambridge University Press, 2005.

\bibitem{flop_13}
K.~Ntontin, M.~D. Renzo, A.~I. Pérez-Neira, and C.~Verikoukis, ``A
  low-complexity method for antenna selection in spatial modulation systems,''
  \emph{IEEE Communications Letters}, vol.~17, no.~12, pp. 2312--2315, December
  2013.

\bibitem{large_no_02}
E.~Biglieri, G.~Taricco, and A.~Tulino, ``Performance of space-time codes for a
  large number of antennas,'' \emph{IEEE Transactions on Information Theory},
  vol.~48, no.~7, pp. 1794--1803, Jul 2002.

\end{thebibliography}

\end{document}